\documentclass[12pt,notitlepage]{revtex4}
\usepackage{graphicx}
\usepackage{amsmath}
\usepackage{amsfonts}
\usepackage{amssymb}
\usepackage{epstopdf}

\font\msytw=msbm9 scaled\magstep1
\font\msytww=msbm7 scaled\magstep1

\let\a=\alpha   \let\g=\gamma  \let\d=\delta \let\e=\varepsilon
      \let\k=\kappa \let\l=\lambda
\let\m=\mu        \let\x=\xi     \let\p=\pi    \let\r=\rho
\let\s=\sigma \let\t=\tau    \let\c=\chi
   
\let\G=\Gamma \let\D=\Delta  \let\L=\Lambda

\def\DD{{\cal D}}

   \def\pp{{\bf p}}
 \def\xx{{\bf x}} \def\yy{{\bf y}} 

\def\kk{{\bf k}}
\def\Vn{{\bf n}}

\def\RRR{\hbox{\msytw R}} \def\rrrr{\hbox{\msytww R}}
 \def\CCC{\hbox{\msytw C}}

 \def\ZZZ{\hbox{\msytw Z}}
\def\zzzz{\hbox{\msytww Z}} 

\def\ss{{\underline{\sigma}}}

\def\\{\hfill\break} 
\let\==\equiv
\let\io=\infty 
\let\0=\noindent

\def\sign{{\rm sign}}
\def\const{{\rm const}}
\def\tende#1{\,\vtop{\ialign{##\crcr\rightarrowfill\crcr\noalign{\kern-1pt
    \nointerlineskip} \hskip3.pt${\scriptstyle #1}$\hskip3.pt\crcr}}\,}
\def\otto{\,{\kern-1.truept\leftarrow\kern-5.truept\to\kern-1.truept}\,}

\def\to{\rightarrow}

\def\qed{\hfill\raise1pt\hbox{\vrule height5pt width5pt depth0pt}}

\def\be{\begin{equation}}
\def\ee{\end{equation}}
\def\bea{\begin{eqnarray}}
\def\eea{\end{eqnarray}}
\def\nn{\nonumber}

\begin{document}

\title{Checkerboards, stripes and corner energies   in spin models with 
competing  interactions}
\thanks{\copyright\, 2011  by the authors. This paper may be
reproduced, in its entirety, for non-commercial purposes.}
\author{Alessandro Giuliani}
\affiliation{Dipartimento di Matematica, Universit\`a di Roma Tre,
L.go S. L. Murialdo 1, 00146 Roma - Italy}
\author{Joel L. Lebowitz}
\affiliation{Departments of Mathematics and Physics, Rutgers University,
Piscataway, NJ 08854 USA.}
\author{Elliott H. Lieb}
\affiliation{Departments of Physics and Mathematics, 
Princeton University, P.O. Box 708, Princeton, NJ 08542-0708 USA.}
\vspace{3.truecm}
\date{June 2, 2011,   \  \    version corner12}

\begin{abstract} 
  We study the zero temperature phase diagram of Ising spin systems
  in two dimensions in the presence of competing interactions, long
  range antiferromagnetic and nearest neighbor ferromagnetic of
  strength $J$.  We first introduce the notion of a ``corner energy''
  which shows, when the antiferromagnetic interaction decays faster
  than the fourth power of the distance, that a striped state
  is favored with respect to a checkerboard state when $J$ is close 
  to $J_c$, the transition to the ferromagnetic state, i.e., when the length
  scales of the uniformly magnetized domains become large. 
  Next, we perform detailed analytic computations on the
  energies of the striped and checkerboard states in the cases of 
  antiferromagnetic interactions with exponential decay and with power law decay
  $r^{-p}$, $p>2$, that depend on the Manhattan distance instead of the 
  Euclidean distance. We prove that the striped phase is 
  always favored compared to the checkerboard phase when the scale of the 
  ground state structure is very large. This happens for $J\lesssim J_c$ if 
  $p>3$, and for $J$ sufficiently large if $2<p\le 3$. Many of our 
  considerations involving rigorous bounds carry over to dimensions greater 
  than two and to more general short-range ferromagnetic interactions.

\end{abstract}

\maketitle

\renewcommand{\thesection}{\arabic{section}}


\section{Introduction}\label{sec_int}
\setcounter{equation}{0}
\renewcommand{\theequation}{\ref{sec_int}.\arabic{equation}} 
In this paper we continue our study of the ground state (GS) of
lattice spin systems with competing ferro (F) and anti-ferro (AF)
Ising-like spin interactions. 
See \cite{GLL1,GLL2,GLL3,GLL4} for previous results.
Such systems are simplified models of
real systems with both short range attractive interactions and long
range dipolar type interactions.  The competitive nature of these
interactions is believed to be responsible for the formation of
mesoscopic periodic structures, such as stripes, in many quasi
two-dimensional (2D) systems at low temperature, see \cite{
B01,Br,DMW00,DKMO,EFKL00,EK93,GD82,HS,JKS05,LCD01,LEFK94,MWRD95,MS92,Mu94,Mu10,OLD08,RKFFK06,
SLPRVP10, S03, SK04,SK06,SS00} for 
several examples of spontaneous pattern formation in physical
systems with competing interactions. See also \cite{KM10,T52} where such competition
is held responsible for the development of macroscopic patterns in chemical and
biological systems described by reaction-diffusion equations.

While it is simple to understand that the competition between interactions 
acting on different length scales can give rise to mesoscopic structures, it is
very difficult to predict the optimal shape of these structures. 
Here we show for a large class of interactions that stripes are 
energetically favorable as compared to other natural structures, 
such as rectangular or square checkerboard.

The Hamiltonians we consider have the form
\be H=\frac{1}2\sum_{\xx\neq \yy}\big[-J\,\d_{|\xx-\yy|,1}+\e\, v(\xx-\yy)\big]
\big(\s_\xx\s_\yy-1\big)\=\frac{1}2\sum_{\xx\neq \yy}\phi(\xx-\yy)
\big(\s_\xx\s_\yy-1\big)\;,\label{1.1}\ee
where $\xx\in\ZZZ^d$, $\s_\xx=\pm1$ are Ising spins, $J$ and $\e$ are two 
positive constants (the strengths of the F and AF interactions), $v$ is a 
non-negative potential, symmetric with respect to $90^o$ rotations and 
summable. In the following, we will be mostly concerned with $d=2$
and $v$ of infinite range. The constants $J$ and $\e$ will
be thought of as being ``large" and ``small", respectively.

The goal is to understand the zero temperature phase diagram as the
ratio $J/\e$ is varied. If $\e=0$, then the ground state is
ferromagnetic. In the opposite limit, that is, $J=0$, then the
ground state displays some non-trivial alternation between positively
and negatively magnetized spins; e.g., if $v(\xx)=|\xx|^{-p}$, $p>d$,
then the ground state is the period-2
antiferromagnetic N\'eel state \cite{FILS}. As the ratio $J/\e$ is
increased from zero to large values, the GS changes
to reduce the number of antiferromagnetic bonds,
presumably by displaying mesoscopic uniformly magnetized structures of larger and
larger  lengths.  It is often assumed that the ground state
configurations are periodic, and display either
checkerboard or striped order, depending on the specific choice of the
interaction and the value of $J$.  In \cite{MWRD95}, it was shown that
for $v(\xx)=|\xx|^{-3}$ and $J$ large enough, the optimal striped
configuration has lower energy than the optimal checkerboard
configuration. This leads to the conjecture (still unproven) that the
ground state configurations of Eq.(\ref{1.1}) with $v(\xx)=|\xx|^{-3}$
and $J$ large display periodic striped order. 

There is in fact evidence for the fact that the sequence of transitions
to the ferromagnetic phase has some universal features \cite{EK93,JKS05,S03,SK04} 
and that the emergence of stripes is essentially {\it independent of the details of
the F and AF interactions}. However, the reason for this is still
unclear and puzzling because stripes break the symmetry of the lattice.

In this paper, we prove that striped patterns
are favored, within a natural class of variational states, when the scale of 
the GS structure is very large compared to the range of the FM interaction. 
A simple explanation of this fact can be based on the concept of {\it corner energy}, which 
suggests that the intersection points among straight phase separation lines can
be thought of as elementary excitations of the system with positive energy, at 
least in the case that the AF interaction decays faster then $r^{-4}$ at large 
distances. Our argument is substantiated by explicit computations in the simple
case that the AF interaction depends on the Manhattan ($L^1$) distance between 
sites and decays as $r^{-p}$, $p>2$, at large distances.

The rest of the paper is organized as follows. In Section \ref{sec_des}
we introduce the notions of line and corner energies and present our argument
explaining why stripes are favored as compared to checkerboard when the AF
interaction decays at infinity faster than $r^{-4}$ and the scale becomes very 
large compared to the lattice spacing. In Section \ref{sec2} 
we present detailed analytical computations of the stripe and checkerboard 
energies in cases where the AF interaction depends on the Manhattan distance 
between sites and decays either exponentially or as a power law $r^{-p}$, $p>2$. 
In Appendix \ref{app1}, we rigorously
compute the critical strength $J_c$ of the FM interaction separating a FM
from a non FM phase, when the AF interaction decays at infinity faster than 
$r^{-3}$. In Appendix \ref{appB}, we prove that power law interactions depending
on the Euclidean distance between sites are reflection positive. This implies that 
if the GS consists of stripes it will be periodic. Finally, in 
Appendix \ref{app3} we discuss in some more detail the zero temperature phase
diagram of the model when the AF potential is an exponential Kac interaction: 
in this case, we have evidence for a transition from checkerboard to stripes as $J$ is increased from zero to $J_c$. The conjecture is verified
by rigorous upper and lower bounds on the GS energy. 

\section{Lines and corners}\label{sec_des}
\setcounter{equation}{0}
\renewcommand{\theequation}{\ref{sec_des}.\arabic{equation}}

In this section, we show that the formation of stripes of mesoscopic size in $d=2$ is essentially 
independent of the nature of the AF interaction in Eq.(\ref{1.1}), as long as it is long range 
and falls off  faster than $|\xx|^{-4}$, i.e., $0\le v(\xx)\le K|\xx|^{-4-\d}$ 
for some constants $K,\d>0$. According to this argument, the 
occurrence of stripes is related to the sign and the relative sizes of  line
and corner energies, which we now define.  

Consider a system in a square box of side length $L$ with half 
the spins up and half down, separated by a vertical line, called an anti-phase 
boundary. When the falloff of $v$ is faster than $|\xx|^{-3}$ 
the energy divided by $L$ will have a nice limit as $L\to\infty$, which is defined to be the 
{\bf line energy} $\t$:
\begin{equation}
\t =  -2 \lim_{L\to \infty} L^{-1}\sum_{\substack{-L/2< x_1 \leq 0\\1\leq 
x_2 \leq L}} \ \sum_{\substack{1\le y_1 \leq L/2 \\1\leq y_2 \leq L}} 
\phi (\xx -\yy).\end{equation}
The energy per unit length $\t$ has the interpretation of {\it surface 
tension} of an infinite straight line, and is linear in $J$, i.e., 
$\t=2(J-J_c)$ for a suitable positive constant $J_c$.

At $J=J_c$, the surface tension of an infinite straight line vanishes
and there is coexistence of the FM 
ground state with the ground state corresponding to a single isolated
anti-phase boundary. It is intuitive that for all $J> J_c$, the ground state is ferromagnetic, since the 
energies of ferromagnetic contours (or, at least, of {\it straight} FM 
contours) is positive. See Appendix \ref{app1} for a proof of stability of the 
FM state against arbitrary contours. For $J<J_c$ the GS is certainly not FM, because the system
reduces its energy by producing anti-phase boundaries. 

Next, we define a {\bf corner energy}, $\k$, by first taking two
crossed, vertical and horizontal,
anti-phase boundaries in the box of size $L$. The energy of
this configuration is, to first approximation, $2\t L$. The
difference between this energy and $2\t L$ has a limit as $L\to
\infty$ whenever the falloff of $v$ is faster than $|\xx|^{-4}$.
This difference is the corner energy $\k$,  and is given by the formula
\begin{equation} \label{one}
\k = 4\sum_{\xx \in Q_1} \sum_{\yy\in Q_3} \phi(\xx-\yy)
+4\sum_{\xx \in Q_2} \sum_{\yy\in Q_4} \phi(\xx-\yy) ,\end{equation}
where $Q_1,Q_2,Q_3,Q_4$ are the first, second, third and fourth quadrant in 
$\ZZZ^2$, respectively.   Note that $\k$ does not depend on the nearest 
neighbor interaction energy and is therefore  positive for the 
Hamiltonian in Eq.(\ref{1.1}).

We now observe that if the GS is made up of rectangles then it necessarily 
consists of a mixture of horizontal and vertical lines, and hence has corners 
where these lines intersect. To lower the energy one can replace the horizontal
lines by the  same number of vetical lines, thereby eliminating the corners. 
While the increased density of vertical lines increases the energy, the saving 
on the corners more than makes up for it when the scale is large enough and 
$J\lesssim J_c$. In fact, consider a configuration of sparse 
straight lines, all at a mutual distance larger than $R\gg1$. 
The interaction energy of any vertical (resp. horizontal) line
in a square box of side $L$ with all the other vertical (resp. horizontal)
lines is positive and smaller than $(\const.)\,L\,R^{-1-\d}$, which follows from
the fact that $0\le v(\xx)\le K|\xx|^{-4-\d}$. Similarly, the interaction energy
of any corner with all the other corners is negative and smaller in absolute 
value than $(\const.)\,R^{-\d}$. Therefore, the total energy $E_\L$ 
of a configuration of widely separated straight lines in a square box 
$\L\subset \ZZZ^2$ of side $L$ has the form
\begin{equation}
E_\L=\big(\t+O(R^{-1-\d})\big)(M_1+M_2) L +\big(\k-O(R^{-\d})\big)\, M_1M_2\; ,
\label{energy}\end{equation}
where $M_1$ ($M_2$) is the total number of horizontal (vertical) lines. 
Eq.(\ref{energy}) shows that, for given $M=M_1+M_2$ of order $L$, it is 
energetically favorable to have $M_1M_2=o(L^2)$. In fact, if the number of corners had a 
finite density, than we could decrease
the energy by rotating all the vertical (horizontal) lines by 90$^o$, 
making them horizontal (vertical)
and placing them half-way between the existing horizontal (vertical) lines. 
After this flipping, the final configuration would have an energy equal to 
$\big(\t+O(R^{-1-\d})\big)(M_1+M_2) L$, which is strictly smaller than 
the one of the initial configuration. 

{\it In this sense, corners play the role of elementary excitations, with a 
positive energy cost, which can be eliminated by rotating straight lines by 
90$^o$.}
A similar analysis shows that also the ``half corners'' produced each time that 
a non-straight anti-phase boundary has a 90$^o$ turn have a finite
positive cost. We are, however, not able to exclude the presence of more complicated
``excitations'' in the GS. 

Regarding the condition on the large distance decay of the AF interactions, 
we do not think it is sharp. However, in the general case, 
the balance between the corner and line energies is much more subtle. 
In fact, if the decay of the AF potential is $\sim r^{-p}$, $2<p< 4$, 
then the corner energy is formally infinite; 
however, corner-corner interactions have an oscillatory 
sign and such oscillations make the effective energy of each corner 
finite and approximately proportional to $R^{4-p}$ if $2<p<4$,
where $R$ is the distance to the neighboring corner. 
It is straightforward to check that if the corners have a finite density, then 
their contribution to the specific GS energy is comparable to the line-line 
interaction and of the order of $R^{2-p}$, where $R$ is the typical separation 
between lines. Therefore, by rotating the vertical lines by 90$^o$, 
we gain the corner energies and lose some line-line interaction energy, both 
of the order $(\const.)R^{2-p}L^2$; 
to decide whether the saving makes up for the loss, we need to compute 
the constant prefactors. This will 
be done analytically in the next section, in the special case of 
AF interactions depending on the Manhattan distance between sites. 
The computation shows that when we rotate the vertical 
lines by 90$^o$ and eliminate the corners, the saving overcomes the loss for all $p>2$. It 
remains to be seen whether this saving is an accident of the specific  
model considered below or whether there is a general physical reason 
behind the result. 

We note that in the special case that the AF interaction is reflection positive
\cite{FILS}, given that the configurations entering the GS are all straight 
vertical (horizontal) lines, then they have to be periodically arranged. 
This follows from the analysis in \cite{GLL1,GLL2,GLL3,GLL4}. 

\section{Comparison of the stripe and checkerboard 
energies}\label{sec2} 
\setcounter{equation}{0}
\renewcommand{\theequation}{\ref{sec2}.\arabic{equation}}

In this section we perform explicit analytic computations of the energies
of the stripe and checkerboard states, for different choices of the fall off of the long range 
AF potential. We focus on the (analytically)
simple case of interactions depending on the Manhattan ($L_1$)
distance $\|\xx\|_1:=|x_1|+|x_2|$ between sites. Our calculations complement
and simplify those in \cite{MWRD95}. 

Let us consider Eq.(\ref{1.1}) with $d=2$, $\e=1$ and 
\begin{equation}
v(\xx)=\int_0^\io d\a\, \m(\a) e^{-\a\|\xx\|_1}\;,\label{1.1a0}
\end{equation}
with $\m(\a)$ a positive measure. We will be particularly concerned with two 
cases:
\begin{enumerate}
\item {\it Exponential interactions}, $v(\xx)=\g^2e^{-\g\|\xx\|_1}$, 
corresponding to the choice $\m(\a)=\g^2  \d(\a-\g)$ in Eq.(\ref{1.1a0});
\item {\it Power law interactions}, $v(\xx)=\|\xx\|_1^{-p}$, with $p>2$,
corresponding to the choice $\m(\a)=\a^{p-1}/\G(p)$ in Eq.(\ref{1.1a0}). 
\end{enumerate}
As mentioned above, the choice Eq.(\ref{1.1a0}) is made to simplify the 
computations; choosing euclidean rather than Manhattan distance should not make
a difference from the physical point of view. Let us remark that the potential 
in Eq.(\ref{1.1a0}) is {\it reflection positive} \cite{FILS} and so is 
the (more usual) power law potential $v(\xx)=|\xx|^{-p}$, with $|\xx|=
\sqrt{x_1^2+x_2^2}$ the Euclidean distance (see Appendix \ref{appB}). 
The property of reflection positivity is not explicitly used in the computations below but, as 
observed at the end of previous section, it implies that if the GS consists of stripes, then these must
be regularly spaced, see \cite{GLL1,GLL2,GLL3,GLL4}. 

Let $s_h(x)$ be the 1D profile of period $2h$, obtained by extending 
periodically over $\ZZZ$ the function $f: (-h,h]\to\RRR$ such that 
$f(x)=\sign(x-1/2)$ for $x=-h+1,\ldots,h$. Let $e_{c}(h)$ be  
the specific energy of the checkerboard configuration, 
$\s_\xx=s_h(x_1)s_h(x_2)$, let $e_s(h)$ be the specific energy of the
striped configuration $\s_\xx=s_h(x_1)$. 
We start by computing the specific energy 
$e(h_1,h_2)$ of the ``rectangular'' configuration $s_{h_1}(x_1)s_{h_2}(x_2)$.
We have:
\bea && e(h_1,h_2)=\label{2.2}\\
&&=\frac{2J}{h_1}+\frac{2J}{h_2}-\frac{1}{h_1 h_2}
\int_0^\io d\a\,\m(\a)\hskip-.2truecm
\sum_{\substack{1\le x_1\le h_1\\ 1\le x_2\le h_2}}\,\sum_{\yy\in\zzzz^2}
e^{-\a|x_1-y_1|}e^{-\a|x_2-y_2|}
\c(\s_\xx\neq\s_\yy)\;,\nn\eea
where $\c({\rm condition})$ is $=1$ if the condition is satisfied, and $=0$
otherwise. 
After some straightforward algebra,
\bea && e(h_1,h_2)=\frac{2 J}{h_1}+\frac{2 J}{h_2}+2\int_0^\io\frac{d\a}{\a^2}
\,\m(\a)\cdot\label{2.4}\\
&&\cdot\Big[-A_\a\frac{\tanh(\a h_1/2)}
{\a h_1/2}-A_\a \frac{\tanh (\a h_2/2)}
{\a h_2/2}+B_\a \frac{\tanh (\a h_1/2)}
{\a h_1/2}\frac{\tanh (\a h_2/2)}
{\a h_2/2}\Big]\;,\nn\eea
where 
\be A_\a=\frac{
(\a/2)^3\cosh(\a/2)}{\sinh^3(\a/2)}
\;,\qquad\quad B_\a=\frac{
(\a/2)^4}{\sinh^4(\a/2)}\;.\label{3.2}\ee
Note that, for small $\a$, $A_\a\simeq 1-(1/15)(\a/2)^4$ and $B_\a\simeq
1-(2/3)(\a/2)^2$, which will be useful in the following.

Using Eq.(\ref{2.4}), we see that 
the energy of a striped configuration of period $h$ is equal to
\be e_s(h/2)=\frac{4J}{h}+2
\int_0^\io\frac{d\a}{\a^2}\,\m(\a)\Big[-2A_\a\frac{\tanh(\a h/4)}{\a h/2}\Big]
\;,\label{e_s}\ee
while the one of a checkerboard configuration of period $2h$ is
\be e_c(h)=\frac{4J}{h}+2
\int_0^\io\frac{d\a}{\a^2}\,\m(\a)\Big[-2A_\a\frac{\tanh(\a h/2)}{\a h/2}+B_\a
\frac{\tanh^2(\a h/2)}{(\a h/2)^2}\Big]\;.\label{e_c}\ee
It is interesting to note that the various terms in 
Eqs.(\ref{2.4})-(\ref{e_s})-(\ref{e_c}) have a clear interpretation in terms
of the notions of ``line energy'' and ``corner energy'', introduced in Section 
\ref{sec_des}. In fact, looking at Eq.(\ref{2.4}),
the terms proportional to $J$ correspond to the FM surface tension energy;
the integral terms with the integrand proportional to $A_\a$ 
correspond to the AF line energy (including both the negative AF surface 
tension and the repulsive line-line interactions); 
the integral term with the integrand proportional to $B_\a$ 
corresponds to the AF corner energy (including both the positive 
corner self-energy and the attractive corner-corner interactions).
The analogous terms in Eqs.(\ref{e_s})-(\ref{e_c}) have a similar 
intepretation; note that $e_c(h)$ includes a positive contribution from 
the corner energy, which does not appear in $e_s(h/2)$, while the 
contribution from the line energy is smaller than the corresponding one in 
$e_s(h/2)$. 

As we discussed above, the goal is to find the balance between these
terms when the scale of the relevant structures is large compared to the 
lattice spacing. We will in fact show that when 
$h\gg 1$, then $e_c(h)>e_s(h/2)$, which is equivalent to 
\be \frac12\int_0^\io\frac{d\a}{\a^2}\,\m(\a)\,B_\a
\frac{\tanh^2(\a h/2)}{(\a h/2)^2}>
\int_0^\io\frac{d\a}{\a^2}\,\m(\a)
\,A_\a\frac{\tanh(\a h/2)-\tanh(\a h/4)}{\a h/2}\,
\label{ec>es}\ee
implying that the GS is striped. This will be proved below by treating 
separately the cases of exponential interactions and of power law interactions,
with $p>4$, $p=4$, $3<p<4$, $p=3$, $2<p<3$ (that are listed here in the order
of increasing difficulty).

{\bf Remark.} Even though Eq.(\ref{ec>es}) does not involve the parameter
$J$, the condition that the scale $h$ of the GS structures is large compared to
the lattice spacing is satisfied only if $J$ is chosen properly. More 
precisely,  as discussed in Section \ref{sec_des} (see also Appendix 
\ref{app1}), if the AF interaction decays faster than $r^{-3}$, then there 
exists a finite $J_c$ such that the homogenous FM state is the GS for all 
$J\ge J_c$; in this case, the condition that $h\gg1$ is valid in the range 
$J\lesssim J_c$. On the contrary, if the decay of the AF interaction is equal 
to $r^{-3}$ or slower, then the condition $h\gg1$ is verified for all $J\gg1$. 
The results below are relevant for $J$ belonging to these ranges. \\

\subsection{Exponential interactions and power laws with $p>4$}

If the AF interaction decays exponentially or as a power law with $p>4$, 
then we already know from the analysis in Section \ref{sec_des} that 
$e_c(h)>e_s(h/2)$ for all $h\gg1$. For completeness, let us check this 
analytically, using Eq.(\ref{ec>es}). In the case of exponential interactions, 
the condition reduces to 
\be \frac12\,B_\g
\frac{\tanh^2(\g h/2)}{(\g h/2)^2}>
\,A_\g\frac{\tanh(\g h/2)-\tanh(\g h/4)}{\g h/2}\,
\label{expec>es}\ee
which is obviously satisfied for $h$ large, simply because the 
l.h.s. goes to zero as $h^{-2}$, while the r.h.s. goes to zero exponentially 
fast in $h$. In the case of power law interactions with $p>4$, 
the l.h.s. of Eq.(\ref{ec>es}) can be rewritten as
\be \frac12\int_0^\io\,d\a\,\a^{p-3}\,B_\a
\frac{\tanh^2(\a h/2)}{(\a h/2)^2}=\frac2{h^2}
\int_0^\io\,d\a\,\a^{p-5}\,B_\a+O(\frac1{h^{p-2}})\;,\label{p4.1}\ee
while the r.h.s. is
\bea && 
\int_0^\io\,d\a\,\a^{p-3}\,A_\a\frac{\tanh(\a h/2)-\tanh(\a h/4)}{\a h/2}=\nn\\
&&=\frac1{h^{p-2}}\int_0^\io\,d\a\,\a^{p-3}\,\frac{\tanh(\a /2)-\tanh(\a /4)}
{\a /2}+O(\frac1{h^{p+2}})\;,\label{p4.2}\eea
where in estimating the error term of order $h^{-p-2}$ we used the fact that 
$|A_\a-1|\le C\a^4$, for a suitable constant $C$. Therefore, 
Eq.(\ref{ec>es}) is valid, simply because $h^{-2}\gg h^{-p+2}$, $\forall p>4$,
for $h$ large. 

\subsection{The case $p=4$}

This case is very similar to the previous one. In fact, the r.h.s. can be 
rewritten and estimated exactly as in Eq.(\ref{p4.2}), with $p=4$; in 
particular, it is $\sim h^{-2}$. The l.h.s. can be rewritten as
\bea  \frac12\int_0^\io\,d\a\,\a\,B_\a
\frac{\tanh^2(\a h/2)}{(\a h/2)^2}&=& \frac2{h^2}\int_{1/h}^1\,\frac{d\a}{\a}\,
\tanh^2(\a h/2)+O(\frac1{h^2})=\nn\\
&=& 2\frac{\log h}{h^2}+O(\frac1{h^2})\;.\label{p4.4}\eea
Therefore,  Eq.(\ref{ec>es}) is valid, simply because $h^{-2}\log h\gg h^{-2}$, 
for $h$ large.

\subsection{The case $p<4$}

\begin{figure}[ht]
\hspace{1 cm}
\includegraphics[height=6.3truecm]{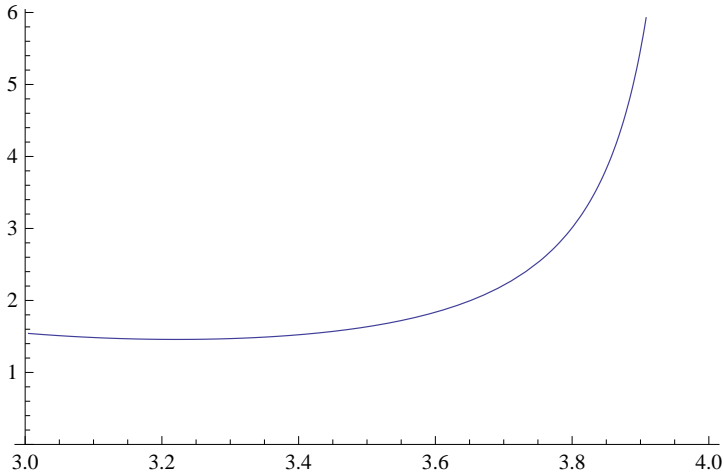}
\caption{A plot of the difference between the right and left hand sides 
of (\protect\ref{3.36}) 
vs $p$, which proves that $e_s(h/2)<e_c(h)$ for all $3<p<4$,
and $h$ large enough.}\label{graph2}
\end{figure}

If $p<4$ the proof of Eq.(\ref{ec>es}) is slightly more subtle, because both 
sides of the inequality scale in the same way as $h\to \infty$. In fact, the 
l.h.s. can be rewritten as
\be  \frac12\int_0^\io\,d\a\,\a^{p-3}\,B_\a
\frac{\tanh^2(\a h/2)}{(\a h/2)^2}=\frac12\Big(\frac2{h}\Big)^{p-2} 
\int_0^\io\,d\a\,\a^{p-3}\,
\frac{\tanh^2\a}{\a^2}+O(\frac1{h^2})\;,
\label{p3.1}\ee
while the r.h.s. reads
\bea&& \int_0^\io\,d\a\,\a^{p-3}\,A_\a\frac{\tanh(\a h/2)-
\tanh(\a h/4)}{\a h/2}=\nn\\ 
&&=
\Big(\frac2{h}\Big)^{p-2} \int_0^\io\,d\a\,\a^{p-3}\,
\frac{\tanh\a -\tanh(\a/2)}{\a}+O(\frac1{h^{p+2}})\;.\label{p3.2}\eea
Therefore,  both sides of Eq.(\ref{ec>es}) scale as $\sim h^{2-p}$ as $h\to\io$. 
The inequality is asymptotically valid if and only if the following condition 
is true:
\be \frac12\int_0^\io\,d\a\,\a^{p-3}\,\frac{\tanh^2\a}{\a^2}>\int_0^\io\,d\a\,
\a^{p-3}\,\frac{\tanh\a -\tanh(\a/2)}{\a}\;.\label{p3.3}\ee
This inequality can be checked numerically in the different ranges $3<p<4$, 
$p=3$ and $2<p<3$. In fact, if $p=3$, Eq.(\ref{p3.3}) is equivalent to
\be \frac12\int_0^\io d\a\,\frac{\tanh^2\a}{\a^2}=0.85256\ldots>
\log 2=0.69315\ldots\label{3.15}\ee
If $3<p<4$, Eq.(\ref{p3.3}) is equivalent to
\be \frac12\int_0^\io d\a\,
\a^{p-5}\tanh^2\a> (2^{p-3}-1)\int_0^\io d\a\,\a^{p-4}(1-\tanh\a)\;.
\label{3.36}\ee
In the limiting case $p\to 3^+$, Eq.(\ref{3.36}) is equivalent to 
Eq.(\ref{3.15}), as it should. In the limit $p\to 4^-$, Eq.(\ref{3.36}) is 
obviously valid (because the r.h.s. tends to a constant, while the l.h.s. 
diverges to $+\io$).
The validity of Eq.(\ref{3.36}) for all values of $p$ in the interval $(3,4)$ 
can be checked numerically, see Fig.\ref{graph2}.

\begin{figure}[ht]
\hspace{1 cm}
\includegraphics[height=7truecm]{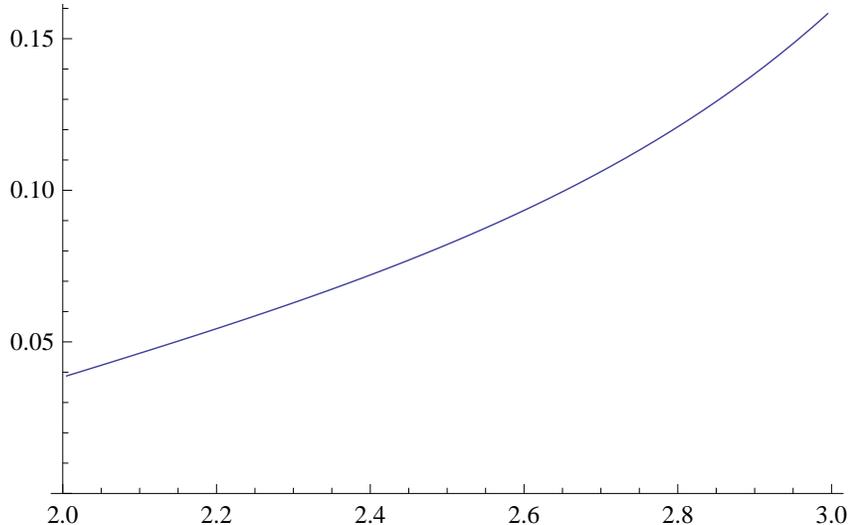}
\caption{A plot of the difference between the right and left hand sides 
of (\protect\ref{3.17c}) 
vs $p$, which proves that $e_s(h/2)<e_c(h)$ for all $2<p<3$,
and $h$ large enough.}\label{graph1}
\end{figure}

Finally, if $2<p<3$, Eq.(\ref{p3.3}) is equivalent to
\be \frac12\int_0^\io d\a\,
\a^{p-5}\tanh^2\a> (1-2^{p-3})\int_0^\io d\a\,\a^{p-4}\tanh\a\;.\label{3.17c}\ee
In the limit $p\to 3^-$, condition Eq.(\ref{3.17c}) reduces to Eq.(\ref{3.15}),
as it should. In the limit $p\to 2^+$, condition Eq.(\ref{3.17c}) reduces to 
\bea&& \log 2>\int_0^\io d\a\Big(\frac{\tanh\a}{\a^2}-\frac{\tanh^2\a}{\a^3}
\Big)\quad\Longleftrightarrow\label{3.18}\\
&& \quad\Longleftrightarrow\quad 
\log2+\frac12=1.193147\ldots>\int_0^\io d\a\frac{\tanh^3\a}{\a^2}=1.154785
\ldots\nn\eea
The validity of Eq.(\ref{3.17c}) for all values of $p$ 
in the interval $(2,3)$ can be checked numerically, see 
Fig.\ref{graph1}. 

This  concludes the proof that $e_c(h)>e_s(h/2)$ whenever $h$ is large, 
for all power law decays with exponent $p>2$ and for exponential 
interactions. An immediate consequence of this analysis is the following: 
let $e^*_s(J)=\min_{h\in\mathbb{N}}e_s(h)$ and $e^*_c(J)=
\min_{h\in\mathbb{N}}e_c(h)$ be the optimal stripe and checkerboard energies at 
a given $J$; then, if the AF interaction is either exponential or power law 
with $p>3$, we have $e^*_s(J)<e^*_c(J)$ for all $J_c-J$ positive and small 
enough; if the AF interaction is power law with $2<p\le 3$, then 
$e^*_s(J)<e^*_c(J)$ for all $J$ large enough. \\

In conclusion, we showed for a 2D spin model with competing short range (nearest neighbor)
FM and long range AF interactions that stripes are favored with respect to checkerboard
when the GS structures are large compared to the range of the FM interaction.
If the AF interaction decays faster than $r^{-4}$, the emergence of stripes close to the 
transition to the FM phase can be understood on the basis 
of a comparison between the sign and relative sizes of the corner and line energies,
which is independent of the details of the AF interaction.  If the decay at infinity 
of the AF interaction is slower, than the balance between corner and line energies is more subtle,
and the understanding of why stripes are favored relies on explicit computations
of the stripe and checkerboard energies, which have been performed here in the simple case that 
the  AF depends on the Manhattan distance between sites.
We believe that future progress on 
the problem will come from a deeper understanding of the reason that interactions that fall off 
slower than $r^{-4}$ always seem to favor stripes.

\acknowledgments We thank O. Penrose and E. Presutti for useful comments and 
conversations. We gratefully acknowledge financial support from the ERC 
Starting Grant CoMBoS-239694 and from PRIN 2008R7SKH2
(A.G.), from NSF DMR-08-02120 and 
AFOSR FA-95550-10-1-0131 (J.L.L.), and from NSF PHY-0965859 (E.H.L.). A.G. thanks
the Institute for Advanced Studies for hospitality during the completion of this work.

\appendix

\section{A rigorous computation of $J_c$}\label{app1}
\setcounter{equation}{0}
\renewcommand{\theequation}{\ref{app1}.\arabic{equation}}

Let us assume that the long range AF interaction decays at infinity faster than $r^{-3}$, and 
let $\t=2(J-J_c)$ be the line energy, as defined in Section \ref{sec_des}. 
Here we want to prove that for all $J\ge J_c$, the homogeneous FM state
is a GS of Eq.(\ref{1.1}) (and is the unique GS for $J>J_c$). 
As already remarked in Section \ref{sec_des}, for $J<J_c$ the homogenoeus state
is {\it not} a GS, simply because the state with a single straight anti-phase 
boundary has negative energy. If $J\ge J_c$ we want to get a lower 
bound on the energy of an arbitrary state, which is positive, unless
the state is homogeneous. 

We proceed in a way similar to the proof of Theorem 3 of \cite{GLL1}. 
We need to introduce some definitions; in particular 
via the basic Peierls construction we introduce the definitions of 
{\it contours} and {\it droplets}.
Given any spin configuration $\ss_\L$ on the squared periodic box $\L$, 
we define $\D$ to be the set of sites at which $\s_i=-1$, i.e., 
$\D=\{i\in\L\,:\,
\s_i=-1\}$. We draw around each $i\in\D$ the $4$ sides of the unit square
centered at $i$ and suppress the sides that occur twice: we obtain 
in this way a {\it closed polygon} $\G(\D)$ which can be thought 
as the boundary of $\D$. Each side of $\G(\D)$ separates a point 
$i\in\D$ from a point $j\not\in\D$. At every vertex of 
$\G(\D)\cap\L^*$, with $\L^*$ the dual lattice of $\L$, there can be either 2 
or 4 sides meeting.
In the case of 4 sides, we deform slightly the polygon, ``chopping off'' 
the edge from the squares containing a $-$ spin.
When this is done
$\G(\D)$ splits into disconnected polygons $\G_1,\ldots,\G_r$ which are called
{\it contours}. Note that, because of the choice of periodic boundary 
conditions, all contours are closed but can possibly wind around the 
box $\L$.
The definition of contours naturally induces a notion of connectedness 
for the spins in $\D$: given $i,j\in\D$ we shall say that $i$ and $j$ 
are connected iff there exists a sequence $(i=i_0,i_1,\ldots,i_n=j)$ such 
that $i_m,i_{m+1}$, $m=0,\ldots,n-1$, are nearest neighbors and none of the 
bonds $(i_m,i_{m+1})$ crosses $\G(\D)$. The maximal connected components 
$\d_i$ of $\D$ will be called {\it droplets} and the set of droplets
of $\D$ will be denoted by $\DD(\D)=\{\d_1,\ldots,\d_s\}$. 
Note that the boundaries $\G(\d_i)$ of the droplets $\d_i\in\DD(\D)$ 
are all distinct subsets of $\G(\D)$ with the property: $\cup_{i=1}^s\G(\d_i)=
\G(\D)$.

Given the definitions above, let us rewrite the energy $E_\L(\ss_\L)$ of 
$\ss_\L$ in a box $\L\subset \mathbb{Z}^2$ with periodic boundary conditions as
\begin{equation}E_\L(\ss_\L)=2J\sum_{\G\in\G(\D)}|\G|-
\sum_{\d\in\DD(\D)}
E_{dip}(\d)\;,\label{2.15}\end{equation}
where
$E_{dip}(\d):= 2\e\sum_{\xx\in\d}\sum_{\yy\in\D^c}v(\xx-\yy)$, which can be bounded
from above as 
\bea E_{dip}(\d)&=&
2\e\sum_{\Vn\in\mathbb{Z}^2}v(\Vn)\sum_{\xx\in\d}\sum_{\yy\in\D^c}
\c(\xx-\yy=\Vn)\le \nn\\
&\le &2\e\sum_{\Vn\in\mathbb{Z}^2}v(\Vn)
\sum_{\xx\in\d}\sum_{\yy\in\mathbb{Z}^2
\setminus\d}\c(\xx-\yy=\Vn)\;.\label{2.17}\eea
Now, the number of ways in which $\Vn=(n_1,n_2)$ 
may occur as the difference $\xx-\yy$ or $\yy-\xx$ with $\xx\in\d$ and 
$\yy\not\in\d$
is at most $\sum_{\G\in\G(\d)}\sum_{i=1}^2|\G|_i|n_i|$, 
where $|\G|_i$ is the number of 
faces in $\G$ orthogonal to the $i$--th coordinate direction.
Therefore, 
\begin{equation}E_{dip}(\d)
\le \e\sum_{\G\in\G(\d)}\sum_{\Vn\in\mathbb{Z}^2}v(\Vn)\sum_{i=1}^2 
|\G|_i|n_i|=2\e\sum_{\G\in\G(\d)}|\G|
\sum_{\substack{\Vn\in\mathbb{Z}^2:\\ n_1>0}}n_1 v(\Vn)=2J_c
\sum_{\G\in\G(\d)}|\G|\;.\label{2.18}\end{equation}
Plugging this back into Eq.(\ref{2.15}) gives
\be E_\L(\ss_\L)\ge2(J-J_c)\sum_{\G\in\G(\D)}|\G|\;,\label{2.20}\ee
which readily implies that the uniformly magnetized state is a GS for all 
$J\ge J_c$ and that it is the only GS for $J>J_c$. 

\section{Reflection positivity of power law interactions}\label{appB}
\setcounter{equation}{0}
\renewcommand{\theequation}{\ref{appB}.\arabic{equation}}

In this Appendix we prove that $v(\xx)=|\xx|^{-p}$, with $p>0$ and 
$|\xx|\=|\xx|_2=\sqrt{x_1^2+x_2^2}$ the usual Euclidean distance, is a 
reflection positive (RP) potential, which may be a useful remark for a possible 
future proof of the periodicity of the GS of Eq.(\ref{1.1}) 
with $v(\xx)=|\xx|^{-p}$. We recall that $v$ is RP if, for all compactly 
supported functions $f:\ZZZ^2\to \CCC$,   
\be \sum_{\substack{x_1,y_1\ge 1\\ x_2,y_2\in\zzzz}}
\bar f_{\xx}\,f_\yy\, v(x_1+y_1-1,x_2-y_2)\ge 0\;.\label{1.1a}\ee
By Schur's product theorem,
pointwise products of RP potentials are reflection positive. 
Therefore, in order to prove that $|\xx|^{-p}$ is RP for all 
$p>0$, it is enough to show that $|\xx|^{-1}$ and $|\xx|^{-\l}$, with $0<\l<1$,
are separately RP. If $v(\xx)=|\xx|^{-1}$ and $x_1\ge 1$,
\bea v(x_1,x_2)&=&\frac1{\sqrt{x_1^2+x_2^2}}=\frac1{2\p^2}\int_{-\io}^{+\io}dk
\int_{-\io}^{+\io}dp
\int_{-\io}^{+\io}dq\,\frac{e^{ikx_1+ipx_2}}{k^2+p^2+q^2}=\nn\\
&=&\frac1{2\p}\int_{-\io}^{+\io}dp
\int_{-\io}^{+\io}dq\,\frac{e^{ipx_2}}{
\sqrt{p^2+q^2}}\,e^{-x_1\sqrt{p^2+q^2}}\;,\label{1.5}\eea
from which (\ref{1.1a}) readily follows. If $v(\xx)=|\xx|^{-\l}$, with 
$0<\l<1$, then (\ref{1.1a}) follows if we prove the stronger result 
\be \int_0^\io d x_1\int_{-\io}^0 d y_1\int_{-\io}^{+\io}d x_2
\int_{-\io}^{+\io}d y_2 \,\frac{\r(\xx)\r(\yy)}{|\xx-\yy|^{\l}}\ge 0
\;,\label{1.6}\ee
if $\r(\xx)$ is a smooth compactly supported real function, with support
contained in $\RRR^2\setminus \{x_1=0\}$, and such that $\r(-x_1,x_2)=
\r(x_1,x_2)$. Using the Fourier transform of $|\xx|^{-\l}$, see e.g.
\cite{LL01} Thm. 5.9, and proceeding as in \cite{FL09}, we can rewrite 
the l.h.s. of (\ref{1.6}) as 
\be\frac{1}{2^\l\p}\frac{\G(1-\frac\l2)}{\G(\frac\l2)} \int_{\rrrr^2}d\kk
\int_{\substack{x_1,y_1>0\\ x_2,y_2\in\rrrr}}d\xx\, d\yy\;\r(\xx)
\frac{e^{i k_1(x_1+y_1)}e^{i k_2
(x_2-y_2)}}{(k_1^2+k_2^2)^{1-\l/2}}\r(\yy)\;.\label{A.7}\ee
We observe that for fixed $x_1+y_1>0$ and $k_2$, the function 
$e^{ik_1(x_1+y_1)}(k_1^2+k_2^2)^{-1+\l/2}$ is analytic in $k_1$ in the upper 
half plane with the cut $\{i\t\,:\,\t\ge |k_2|\}$ removed. Deforming the 
contour of integration in $dk_1$ to this cut and calculating the jump 
of the argument across it we obtain
\be \int_{-\io}^{+\io}dk_1 \frac{e^{ik_1(x_1+y_1)}}{(k_1^2+k_2^2)^{1-\l/2}}
=2\sin\big(\p(1-\l/2)\big)\int_{|k_2|}^\io d\t \frac{e^{-\t(x_1+y_1)}}{
(\t^2-k_2^2)^{1-\l/2}}\;.\label{A.8}\ee
Plugging this back into (\ref{A.7}) we find
\bea && \int_{\substack{-x_1,y_1>0\\ x_2,y_2\in\rrrr}}\,
\frac{\r(\xx)\r(\yy)}{|\xx-\yy|^{\l}}=\frac{2^{1-\l}}{\p}\frac{\G(1-\frac\l2)}
{\G(\frac\l2)} \sin\big(\p(1-
\l/2)\big)\int_{\rrrr}d k_2\int_{|k_2|}^\io d\t\cdot\label{A.9}\\
&&\hskip1cm\cdot\frac1{
(\t^2-k_2^2)^{1-\l/2}}\int_{\substack{x_1,y_1>0\\ x_2,y_2\in\rrrr}}d\xx\,d\yy\;
\Big(\r(\xx) e^{-\t x_1+ik_2x_2}\Big)\,\Big(\r(\yy)e^{-\t y_1-ik_2y_2}\Big)
\;,\nn\eea
which is clearly nonnegative. This concludes the proof that $|\xx|^{-p}$ 
is reflection positive for all $p>0$. 

\section{Kac interactions}\label{app3}
\setcounter{equation}{0}
\renewcommand{\theequation}{\ref{app3}.\arabic{equation}}

In this appendix we add some comments about the possible structure of the GS
of Eq.(\ref{1.1}) in the case that $v$ is a 2D Kac potential, i.e., $v(\xx)=\g^2 v_0(\g\xx)$, 
with $\g$ a small parameter. These may be relevant 
for the understanding of the ``froth problem", addressed by Lebowitz and Penrose in  
\cite{LP66}. To be definite
and make things simple, we restrict to the case of exponential interactions 
depending on the Manhattan distance: $v(\xx)=\g^2 e^{-\g\|\xx\|_1}$. 
In this case, $J_c=2\g^{-1}A_\g$ and, if $J\ge J_c$, the GS is the homogeneous 
FM state. 

From the computations in Section \ref{sec2}, we already know that, 
as $J\to J_c^-$, the stripe state is energetically favored as compared to the 
checkerboard state. If $e^*_s(J)=\min_{h\in\mathbb{N}}e_s(h)$ and $e^*_c(J)=
\min_{h\in\mathbb{N}}e_c(h)$ are the optimal stripe and checkerboard energies,
an explicit computation shows that, if $0<\x:=\g(J_c-J)\ll1$, 
\be  e_s^*(J)= -\frac{2\x}{|\log\x|}+O\Big(\frac{\x\log|\log\x|}{
(\log\x)^2}\Big)\;,\qquad 
e_c^*(J)=-\frac{\x^2}{2 B_\g}+O(e^{-1/\x})\;;\label{bla}\ee
correspondingly, the scales $h^*_s$ and $h^*_c$ of the optimal stripe and 
checkerboard configurations turn out to be:
\be q^*_c:=\frac{\g h_c^*}2=\frac{2B_\g}{\x}+O(e^{-1/\x})\;,\qquad
 q^*_s:=\frac{\g h_s^*}2=\frac12\big|\log\x\big|+O(\log\big|\log\x\big|)\;,
\label{bla2}\ee
Using methods similar to those in the proof of Theorem 3 of \cite{GLL1}, it is 
easy to prove that for $\x$ small the scaling of $e_s^*(J)$ is the optimal one;
i.e., the absolute ground state energy per site $e_0$ admits a lower bound of
the form $e_0\ge -(\const.)\x\cdot|\log\x|^{-1}$.

It is interesting that the model with exponential Kac interactions also displays
a phase where mesoscopic checkerboard are energetically favored with respect to 
stripes. In fact, note that the periods of the optimal 
checkerboard and striped states are given by
\bea && \g J= \Big(2A_\g-2B_\g\frac{\tanh q^*_c}{q^*_c}\Big)
\big[\tanh q^*_c-q^*_c(1-\tanh^2q^*_c)\big]\;,\label{2.600}\\
&& \g J= 2A_\g\big[\tanh q^*_s-q^*_s(1-\tanh^2q^*_s)\big]\;,
\label{2.601}\eea
from which we immediately recognize that, if $\g\ll 1$ and $1\lesssim J\ll 
\g^{-1}$, then $h_c^*$ and $h_s^*$ are both $\ll\g^{-1}$; therefore, the solution 
to these equations can be determined by expanding their r.h.s. in Taylor series
in $q$ and solving to dominant order, which leads to:
\bea && \frac{\g h_c^*}2=\Big(\frac{9\g J}{4}\Big)^{1/5}+O\big((\g J)^{3/5}
\big)\;,\qquad
e_c^*=-2+\frac{10}{9}\Big(\frac{9\g J}{4}\Big)^{4/5}+O\big((\g J)^{6/5}\big)
\;,\nn\\
&& \frac{\g h_s^*}2=\Big(\frac{3\g J}{4}\Big)^{1/3}+O(\g J)\;,\qquad
e_s^*=-2+2\Big(\frac{3\g J}{4}\Big)^{2/3}+O\big((\g J)^{4/3}\big)\;.\nn
\eea
Therefore, in this regime the specific
energy $e_c^*$ of the optimal checkerboard configuration is smaller
than the specific energy $e_s^*$ of the optimal striped configuration. This 
suggests that for any fixed $J$ and $\g$ small enough 
the ground states of the considered model display periodic checkerboard order, 
a conjecture  supported by the fact that the 
absolute ground state energy per site admits a lower bound of the form 
$e_0\ge -2+ (\const.)(\g J)^{4/5}$, which has the right scaling, see below for a
proof. 

\begin{figure}[ht]
\includegraphics[height=8.3truecm]{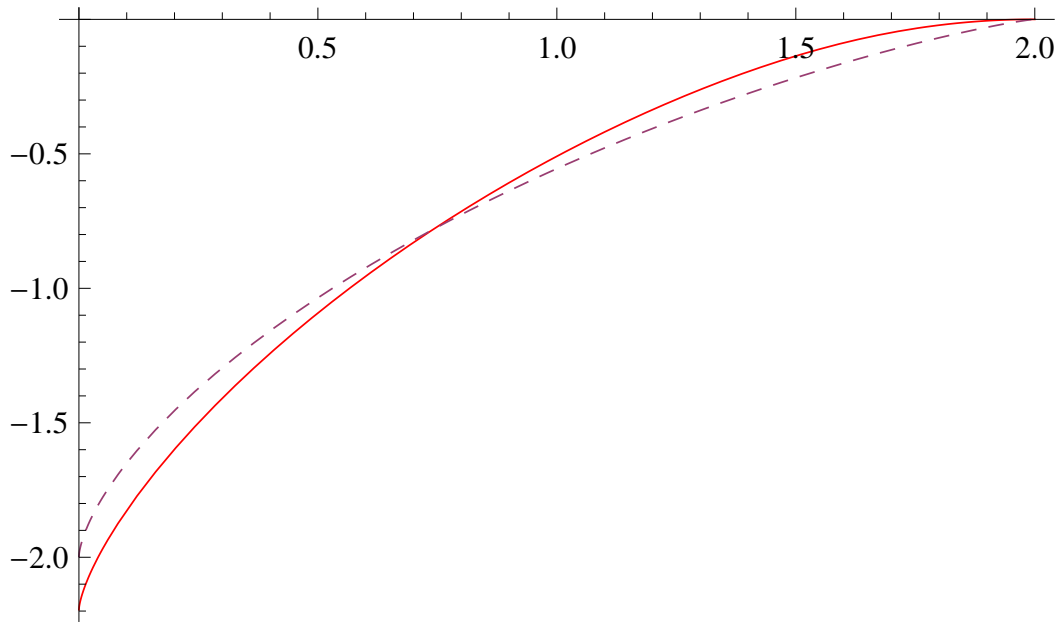}
\caption{A plot of the optimal checkerboard energy $e_c^*$ (solid line) and
of the optimal striped energy $e_s^*$ (dashed line) vs $\tilde J:=\gamma J$ 
for exponential Kac interactions $v(\xx)=\g^2 e^{-\g\|\xx\|_1}$ 
at $\g=0.4$. The plot shows a transition 
from a case where $e_c^*<e_s^*$ (for small values of $\tilde J$) to a case
where $e_c^*>e_s^*$ (for larger values of $\tilde J$). In the limit
$\g\ll1$, the transition is expected to occur for $J$ of the order $\g^{-1}$. 
}\label{??}
\end{figure}

In conclusion, if the AF interaction is exponential with a 
Kac-like scaling, we expect that as $J$ is increased from $0$ to $J_c$, the GS 
should display a transition from checkerboard to stripes. On the basis of the 
previous computations, we expect the trabnsition to take place at values 
of $\g J$ of order one, see Fig.\ref{??}.

{\bf Remark.} The scaling of the checkerboard energy as well as the very 
existence of a checkerboard phase may depend on the specific choice of the Kac 
potential. In particular, it may depend on the reflection positivity 
property of the Kac potential (note that the considered exponential interaction
is reflection positive); if $v_0$ is smoother at the origin (e.g., 
$v_0(\xx)=e^{-|\xx|^2}$), the checkerboard phase may disappear or, at least,
be characterized by a completely different scaling behavior. The reason for 
this is already apparent in a 1D toy model: consider model Eq.(\ref{1.1})
in $d=1$ with $v(x)=\g v_0(\g x)$ and $v_0$ either of the form 
$v_0(x)=e^{-|x|}$ or $v_0(x)=e^{-x^2}$; if one optimizes the energy 
of a configuration consisting of blocks of uniformly magnetized spins of 
size $h$ and alternating sign, the optimal size turns out to be of the order 
$\g^{-2/3}$ in the exponential case and $\g^{-1}/|\log \g|$ in the 
gaussian case. This can be seen as follows: the scale of the optimal periodic 
structure can be found by balancing the energy contributions from the FM and AF
interactions; while the first is $2J/h$, the second is of the order of 
$\hat v_0(1/\g h)$, with $\hat v_0$ the Fourier transform of $v_0$; the latter 
depends on the smoothness properties of $v_0$ and, more specifically, it 
behaves like $\hat v_0(k)\sim k^{-2}$ or $\sim e^{-(\const)k^2}$ at large $k$, 
in the cases of $v_0$ exponential or gaussian, respectively. Minimization of
$2J/h+\hat v_0(1/\g h)$ over $h$ gives the optimal size of the structures. 

The fact that the nature of the checkerboard structure  
depends on the reflection positivity properties of the Kac potential is 
consistent with the fact that the proof of the lower bound on the energy in the 
Kac regime heavily uses reflection positivity, see next subsection.

\subsection{Lower bound on the energy: Kac regime}

Let us assume that $1\lesssim J\ll \g^{-1}$: in this case we want to 
prove that $e_0\ge -2+(\const.)(\g J)^{4/5}$, which asymptotically matches 
the upper bound $e_0\le e^*_c$ and supports the conjecture that, in this
regime, the ground state has checkerboard order. Let $E_\L(\ss_\L)$ be the 
energy of the spin configuration $\ss_\L$ in the periodic squared box $\L$. 
Let us consider a partition of $\L$ into squares $Q_i$ of side $\ell$: 
$\L=\cup_{i=1}^{|\L|/\ell^2}$; given $\ss_\L$ and $Q_i$ we shall denote by 
$\ss_{Q_i}$ the restriction of the spin configuration $\ss_\L$ to the square 
$Q_i$. Let $v_\g^\L(\xx)=\g^2\sum_{\Vn\in\ZZZ^2}e^{-\g|\xx+\Vn L|_1}$ and 
let us rewrite
\be E_\L(\ss_\L)=-2\frac{(\g/2)^2}{\tanh^2(\g/2)}|\L|+
E^\L_\g(\ss_\L)+E^\L_J(\ss_\L)\;,\label{2.701}\ee
where 
$E^\L_\g(\ss_\L)=\frac12\sum_{\xx,\yy\in\L}v_\g^\L(\xx-\yy)\s_\xx\s_\yy$
is the antiferromagnetic energy associated to $\ss_\L$,
while $E^\L_J(\ss_\L)=2J\sum_{\xx\in\L}\sum_{i=1}^2\c(\s_\xx\neq
\s_{\xx+\hat e_i})$ is the surface tension energy of $\ss_\L$ in the box
$\L$ with periodic boundary conditions.  
If we drop the surface tension energy across the boundaries of the squares 
$Q_i$, we get a lower bound on the energy of the form:
\be E_\L(\ss_\L)\ge -2\frac{(\g/2)^2}{\tanh^2(\g/2)}|\L|+
E^\L_\g(\ss_\L)+\sum_{i=1}^{|\L|/\ell^2}\widetilde
E^{Q_i}_J(\ss_{Q_i})\;,\label{2.7}\ee
where $\widetilde E^{Q_i}_J(\ss_{Q_i})$ is the surface tension energy 
of the spin configuration $\ss_{Q_i}$ in the box $Q_i$ with open boundary 
conditions. If $m_i:=\ell^{-2}\sum_{\xx\in Q_i}(\ss_{Q_i})_\xx$, the surface 
tension energy can be further bounded from below by:
\be \widetilde
E^{Q_i}_J(\ss_{Q_i})\ge 2J \ell\min\{1,2\sqrt{2(1-|m_i|)}\}\;.\label{2.8}\ee
Moreover, using reflection positivity \cite{FILS}, the antiferromagnetic energy
can be bounded from below as
\be E^\L_\g(\ss_\L)\ge \ell^2\sum_{i=1}^{|\L|/\ell^2}e_\g(\ss_{Q_i})\;,
\label{2.9}\ee
where $e_\g(\ss_{Q_i})$ is the specific energy of the infinite volume 
configuration obtained by repeatedly reflecting $\ss_{Q_i}$ 
(with ``antiferromagentic reflections'') across the sides of $Q_i$ and its 
images. More explicitly,
\be e_\g(\ss_{Q_i})=\frac2{\ell^4}\!\!\!\!\!\sum_{\substack{
\pp=\frac{\p}{\ell}(n_1,n_2)\\ n_i=1,3,5,\ldots,2\ell-1}}
\!\!\!\!|\widetilde\s_\pp|^2\g^2(1-e^{-2\g})^2\prod_{i=1}^2\frac1{
(1-e^{-\g})^2+2e^{-\g}(1-\cos p_i)}\;,\label{2.10}\ee
where $\widetilde\s_\pp:=\sum_{\xx\in Q_i}\s_\xx e^{-i\kk\xx}$. 
Using the fact that, for all $\e>0$,
\be |\widetilde\s_\pp|^2\ge \frac{4(1-\e)}{(1-\cos p_1)(1-\cos p_2)}-\frac1{\e}
\ell^4(1-|m_i|)^2\;,\label{2.11}\ee
we get 
\be e_\g(\ss_{Q_i})\ge (1-\e) 
e_\g(\underline{1}_{Q_i})-(\const.)\frac1{\e}
(1-|m_i|)^2 (\g \ell)^4\;,\label{2.12}\ee
where $e_\g(\underline{1}_{Q_i})$ is the antiferromagnetic 
energy per site of the checkerboard configuration with tiles of side $\ell$. 
Note that
$e_\g(\underline{1}_{Q_i})$ scales as $(\const.)(\g \ell)^4$ in the regime 
under consideration and for $\ell\gg 1$; moreover, it can be bounded from below
by $\bar C(\g\ell)^4$ for a suitable constant $\bar C$. Optimizing over 
$\e$ we get
\be e_\g(\ss_{Q_i})\ge e_\g(\underline{1}_{Q_i})-c(1-|m_i|) 
(\g \ell)^4\;,\label{2.12a}\ee
for a suitable constant $c$. Combining all the previous bounds we find that 
$E_\L(\ss_\L)+2(\g/2)^2\tanh^{-2}(\g/2)|\L|$ can be 
bounded from below by
\be \ell^2
\sum_{i=1}^{|\L|/\ell^2}\Big\{\frac{2J}{\ell}
\min\{1,2\sqrt{2(1-|m_i|)}\}+\big[\bar C -c
(1-|m_i|)\big](\g\ell)^4\Big\}\;.\label{2.13}\ee
Optimizing over $m_i$ and $\ell$ leads to $\ell=(\const.)(J\g^{-4})^{1/5}$ and 
\be e_0\ge -2\frac{(\g/2)^2}{\tanh^2(\g/2)}|\L|+(\const.)(\g J)^{4/5}\;,
\label{2.14}\ee
as desired. The proof of (\ref{2.14}) can be easily adapted to higher 
dimensions and to cases where the ferromagnetic interaction has finite range
rather than being nearest neighbor. On the contrary, the assumption of RP
was used in a crucial way and it is likely that in the presence of more
general long-ranged antiferromagnetic interactions the ground state energy 
scales differently with $\g$. 


\end{document}